\documentstyle[prc,aps,psfig]{revtex}
\newcommand{\be}{\begin{equation}}
\newcommand{\ee}{\end{equation}}
\newcommand{\bea}{\begin{eqnarray}}
\newcommand{\eea}{\end{eqnarray}}
\newcommand{\non}{\nonumber}
\newcommand{\dn}{{\rm d}}
\newcommand{\re}{{\rm Re}}
\newcommand{\im}{{\rm Im}}
\newcommand{\MeV}{{\rm MeV}}
\newcommand{\fm}{{\rm fm}}

\begin{document}
\baselineskip 16pt
\parindent0em \parskip1.5ex plus0.5ex minus 0.5ex
\title{The nucleon spectral function at finite temperature and the
onset of superfluidity in nuclear matter}
\author{T. Alm, G. R\"opke and A. Schnell}
\address{Arbeitsgruppe der Max-Planck-Gesellschaft
``Theoretische Vielteilchenphysik''
an der Universit\"at Rostock, Universit\"atsplatz 1, 18051
Rostock, Germany}
\author{N.H. Kwong and H.S. K\"ohler}
\address{Physics Department, University of Arizona, Tucson,
AZ 85721, USA}
\maketitle
\begin{abstract}
\begin{sloppypar}
Nucleon selfenergies and  spectral functions are calculated at the
saturation
density of symmetric nuclear matter at finite temperatures.
In particular,  the behaviour of these quantities 
at temperatures above and close to
the critical temperature for the superfluid phase transition in
nuclear matter is discussed.
It is shown how the singularity in the thermodynamic T-matrix
at the critical temperature for superfluidity (Thouless criterion)
reflects in the selfenergy and
correspondingly in the spectral function.
The real part of the on-shell selfenergy (optical potential) shows an
anomalous
behaviour for momenta near the Fermi momentum and temperatures close to the
critical temperature related to the pairing singularity in the imaginary
part. For comparison the selfenergy derived from the K-matrix of Brueckner
theory is also calculated.
It is found, that there is no pairing singularity in the imaginary part
of the selfenergy in this case, which is due to the neglect of hole-hole
scattering in the K-matrix. 
From the selfenergy the spectral function and the occupation numbers
for finite
temperatures are calculated. 
\end{sloppypar}
\end{abstract}
\vspace{2cm}
\hspace*{1.8cm}
(MPG-VT-UR 65/95, submitted to Phys. Rev. {\bf C})
\newpage
\pssilent
\section{Introduction}
Heavy ion reactions at intermediate energies probe the nuclear equation of
state in a broad temperature and density range. One tries to extract
from the observables
signals for cluster formation, multifragmentation or the liquid-gas
phase transition. The theoretical description of such phenomena demands to go
beyond the usual quasiparticle description for the equilibrium properties
(equation of state) as well as for the non-equilibrium properties
(BUU-simulations).
The need to go beyond the quasiparticle approximation in the
description of nuclear
matter is further advocated by the electron scattering experiments
from heavy nuclei.
These experiments give a clear evidence that the one-nucleon spectral function
shows pronounced deviations from mean field estimates, which are due to
NN-correlations \cite{Benhar}. The effect of these correlations can most
accurately be studied for nuclear matter \cite{Benhar1}.

A systematic quantum statistical
approach for the description of dense nuclear matter can be given using
Green function theory. For the description of phenomena like the formation of
clusters (bound states) in the medium or the onset of a superfluid phase
one has to allow for a finite lifetime (damping) of the one particle states.
Such a treatment can be based on the nucleon spectral function,
which is defined in
terms of the real and imaginary parts of the selfenergy. 
Having the nucleon spectral function at the disposal, one is able to determine
the momentum distribution, the response function as well as the
nuclear equation of state including correlations beyond the mean field level.

Most of the existing calculations of the nucleon spectral
function have been done for zero temperature nuclear matter at the
saturation density $n_0$.
In refs. \cite{Baldo1,Koehler1,Vonder1}
it is shown that the
quasiparticle description is only valid near the Fermi energy.
In general, the spectral function has a considerable width
indicating that the one-particle states are strongly damped at
the saturation density.
Baldo et al. \cite{Baldo1} calculated the on- and off-shell
properties of
the nucleon mass operator and the nucleon spectral function.
They stressed the importance to retain the full frequency
dependence of the selfenergy for the calculation of the spectral
function.
K\"ohler \cite{Koehler1} compared the nucleon spectral function calculated 
within Green function theory with a calculation done in Brueckner
theory. He pointed out the near equivalence between the two theories at
zero temperature.
Vonderfecht et al. \cite{Vonder1} used Green function theory to
calculate the nucleon selfenergy within the ladder approximation
at zero temperature. They emphasized the need to correctly
treat the pairing correlations contained in this approximation
if hole-hole propagation is included in the kernel of the vertex function
(thermodynamic T-matrix).
Benhar et al. \cite{ben89} calculated spectral functions
for nuclear matter using the hyper-netted chain (HNC) method.
The spectral function and the nucleon momentum distribution in nuclear
matter were calculated by Benhar et al. \cite{Ben1} at densities below
the saturation density within orthogonal correlated basis functions theory
(OCBF). As a result, they find that with decreasing density the discontinuity
at the Fermi surface decreases. This means that the momentum
distribution does not approach the non-interacting response with
decreasing density. This was interpreted as being due to the attraction
in the N-N interaction leading to the formation
of ``bound pairs'' of nucleons. Van Neck et al. \cite{Neck} use correlated
basis functions theory (CBF)
with the Urbana $v_{14}$ interaction to calculate the nuclear
matter spectral function and momentum distribution below $n_0$.
In agreement with the previous authors they found that the momentum
distribution does not reach the non-interacting one with decreasing
density. 
De Jong et al. \cite{DeJong1} calculated the nucleon spectral
function based on an extension of the relativistic Dirac-Brueckner
scheme.

First finite temperature calculations of the nucleon self energy
were carried out by Grang\'e et al. \cite{Grange1}, R\"opke et al.
\cite{Schulz} (in connection with the e.o.s.) and K\"ohler
\cite{Koehler2}. The latter author found a pronounced temperature
dependence of the spectral function at the saturation density.
Using the extended quasiparticle approximation for the nucleon
spectral function Schmidt et al. \cite{Schmidt} included the formation of
two-nucleon correlations (in particular bound states) in the
equation of state of nuclear matter.
The finite temperature nucleon spectral function in the low-density region
of nuclear matter was calculated by Alm et al. \cite{AlmPLB} 
demonstrating the necessity to include the contribution of
two-nucleon bound states to the one-nucleon spectral function at low density.
These authors find that genuine bound state formation in nuclear matter
is only possible at extremely low densities ($n\leq 0.05\,n_0$,
$T=10\;\MeV$). With increasing density the bound states
are dissolved in the medium (Mott effect). 
However, the formation of two-nucleon pairs in the continuum as well
as their possible  Bose condensation \cite{AlmNP} can take place at
these higher densities as well.
The necessity to include the deuteron singularity in the two-particle
T-matrix into the description of nucleon-nucleus scattering
a low energies within the folding model was demonstrated by Love
et al. \cite{Love}.

Within this paper we would like to discuss the nucleon selfenergy,  
spectral function and momentum distribution at the saturation density and at 
finite temperatures.
In particular, we will concentrate on
low temperatures near the critical temperature
for the superfluid phase transition in symmetric nuclear matter, which has a
maximum of $\sim 5-6\;\MeV$ at densities below $n_0$ \cite{AlmNP}.
This phase transition has recently been discussed by some authors
in particular
with respect to the possible formation of a condensate of neutron-proton-pairs
in the $^3{\rm S}_1-{}^3{\rm D}_1$-channel \cite{Baldo2} and the
relation to the Bose condensation
of deuterons in low-density nuclear matter \cite{Stein}.
Moreover, Baldo et al. \cite{Baldo3} suggest that this could be a possible
mechanism of deuteron formation in heavy-ion reactions at intermediate energy.
The onset of the superfluid phase is contained in the
thermodynamic T-matrix in the ladder approximation \cite{AlmNP}.
It manifests itself as a pole defining the critical temperature
for the thermodynamic T-matrix \cite{Thouless}. It could be shown that already
above the critical temperature the T-matrix shows a
resonance-like behaviour which could be understood as a
precursor effect of the superfluid phase transition
\cite{AlmPRC}.
Consequently, near the critical
temperature the selfenergy and correspondingly the
spectral function  should reflect this resonance-like
behaviour of the T-matrix.
In this paper we will demonstrate how the T-matrix, the nucleon
selfenergy and spectral function are changed when approaching the
critical temperature
for the superfluid phase transition from above.

In chapter II we derive a set of selfconsistent  expressions for
the retarded nucleon selfenergy and the nucleon spectral function at
finite temperature within the framework 
of Matsubara Green functions.
In the following chapter III the differences to the usual Brueckner theory
generalized to finite temperatures are pointed out.
The chapter IV contains results for the T-matrix and the K-matrix
and the corresponding selfenergies including the optical potential
at finite $T$. Within chapter V the
corresponding spectral functions and occupation numbers are presented.
\section{Green Function Formalism and Approximations}
For the derivation of the selfenergy and the spectral function
we use the Matsubara Green function
technique as outlined in ref.\ \cite{Fetter}.
We put $\hbar, k_B=1$ throughout the paper.
From the definition of the one-particle spectral function 
$A(1,\omega)=i [G(1,\omega+i0)-G(1,\omega-i0)]$
and from the retarded selfenergy $\Sigma$ the
spectral function according to the Dyson equation reads 
\be\label{A}
A(1,\omega)=
\frac{2\, \im\,\Sigma(1,\omega)}
{[\omega-\frac{p_1^2}{2m}-\re\,\Sigma(1,\omega)]^2+[\im\,\Sigma(1,\omega)]^2}
\mbox{ ,}\ee
where $1=\{ p_1, \sigma_1, \tau_1\}$ denotes momentum, spin and isospin of a
single particle.\\
The one-particle spectral function (\ref{A}) fulfills the sum rule
\be \label{sum}
\int \frac{\dn\omega}{2 \pi} A(1,\omega)=1
\mbox{ .}\ee
It determines the macroscopic properties of the system such as
the occupation $n_1$ of single particle states given by
\be\label{occ_numb}
n_1(\mu,T)=\int\frac{\dn\omega}{2 \pi} f(\omega) A(1,\omega)\mbox{ ,}
\ee
where $f(\omega)=\{\exp ((\omega-\mu)/T) +1  \}^{-1}$ is the Fermi
distribution.
The equation of state of nuclear matter, where the nucleon density $n$ is a
function of the chemical potential $\mu$ and the temperature $T$, reads
\be\label{n}
n(\mu,T)=\frac{1}{\Omega}\sum_{1}n_1(\mu,T)
\mbox{ ,}\ee
with $\Omega$ being the normalization volume.

For the evaluation of the selfenergy $\Sigma$ in eq.\ (\ref{A})
approximations have to be made. We start from a cluster
decomposition of the selfenergy \cite{GB},
which is given by the following set of diagrams
\begin{figure}
\centerline{\psfig{figure=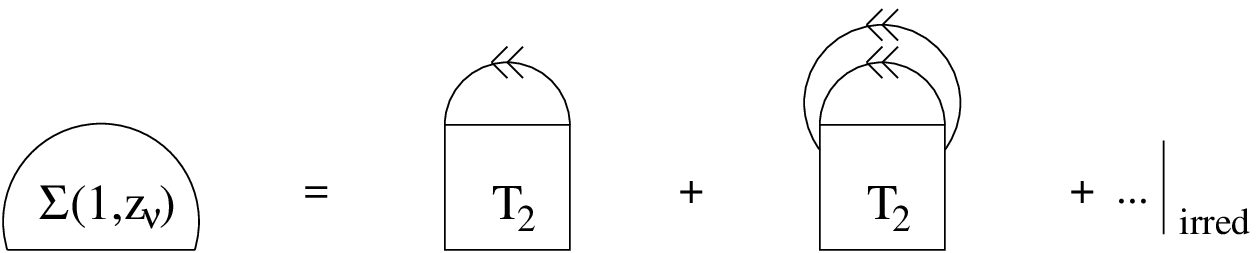,height=2.0cm}\hspace{0.3cm},} 
\end{figure}
\vspace{-1.75cm}
\be\label{clust}\left.\right.\ee
where $z_{\nu}$ is the Fermionic Matsubara frequency.
The cluster decomposition is appropriate
for the consideration of $n$-particle
correlations via the $n$-particle T-matrix $T_n$.
Within this paper we restrict to $n=2$, i.e. to two-particle
correlations represented by the first diagram in eq.\ (\ref{clust}).
Within the ladder approximation the two-particle T-matrix is given by
\be\label{T}
T(121'2',z)=V(121'2')+\frac{1}{\Omega}\sum_{3,4,5,6}
V(1234)\, G_2(3456,z)\,
T(561'2',z)
\mbox{ ,}\ee
where the quantity $G_2$ is defined as the product of two full
one-particle Green functions given in spectral representation as
\be\label{G2}
G_2(121'2',z)=\int \frac{\dn\omega}{2 \pi} \int \frac{\dn\omega'}{2 \pi}
\frac{1-f(\omega)-f(\omega')}{\omega+\omega'-z}
A(1,\omega) A(2,\omega') \delta_{11'} \delta_{22'}
\mbox{ .}\ee
Using the spectral representation for the one particle Green function
as well as for the T-matrix  \cite{GB} the first diagram in (\ref{clust})
yields
\be\label{M2.28}
\Sigma(1,\omega+i0)= \frac{1}{\Omega}\sum_{2}\int\frac{\dn\omega'}{2
\pi}  A(2, \omega')
\left( f(\omega') V_{\rm ex}(1212)-\int \frac{\dn E}{\pi}
\frac{[f(\omega')+g(E)] \im T_{\rm ex}(1212,E)}{E-\omega'-\omega-i0}\right) 
\mbox{ ,}\ee
where $f(\omega)$ is the Fermi distribution and $g(E)=
\{\exp ((E-2\mu)/T)-1\}^{-1}$ 
the Bose distribution function for the two-nucleon states.
$T_{\rm ex}$ and $V_{\rm ex}$ denote matrix
elements of the retarded T-matrix and of the potential including exchange.

The ladder T-matrix equation (\ref{T},\ref{G2}) as well as the
selfenergy (\ref{M2.28}) contain the one-particle spectral function
(\ref{A}) and thus form a selfconsistent set of equations.
A solution of the complicated set of equations (\ref{A}),
(\ref{T},\ref{G2}) and (\ref{M2.28}) can be
achieved by iteration. In the first step of iteration we replace
the spectral function in eq.\ (\ref{M2.28}) by a
quasiparticle spectral function
\be\label{qp}
A^{\rm QP}(1,\omega)=2 \pi\delta(\omega-\epsilon_1)
\mbox{ ,}\ee
where the quasiparticle energy $\epsilon_1$ is defined as
\be\label{eps}
\epsilon_1=\frac{p_1^2}{2 m}+\re\,\Sigma(1,\omega) \mid_{\omega=\epsilon_1}
\mbox{ .}\ee
Inserting the quasiparticle spectral function
(\ref{qp}) in eq.\ (\ref{G2}) results in the quasiparticle approximation
for the two-particle T-matrix (\ref{T}).
Within these approximations the following expressions for the imaginary and
real part of the selfenergy (\ref{M2.28}) are obtained
\be\label{ImSig}
\im\,\Sigma(1,\omega)=\frac{1}{\Omega}\sum_{2}
[f(\epsilon_2)+g(\epsilon_2+\omega)]\:
\im T_{\rm ex}(1212,\epsilon_2+\omega)
\ee
and
\be\label{ReSig}
\re\,\Sigma(1,\omega)= \frac{1}{\Omega}\sum_{2}\left(f(\epsilon_2)\re T_{\rm 
ex}(1212,\epsilon_2+\omega)
-{\bf P}\int\frac{\dn E}{\pi}\frac{g(E)\im T_{\rm 
ex}(1212,E)}{E-\epsilon_2-\omega}
\right)
\mbox{ ,}\ee
where {\bf P} denotes the Cauchy principal value.
$T_{\rm ex}$ ist calculated in the quasiparticle approximation.
Rather then (\ref{ReSig}) we use in the numerical calculation the real
part from the dispersion relation (Kramers-Kronig-relation) in the form
\be\label{DisS}
\re\,\Sigma(1,\omega)= \Sigma^{\rm HF}(1)+{\bf P}\int\frac{d\omega'}{\pi}
\frac{\im\,\Sigma(1,\omega')}{\omega-\omega'}
\mbox{ ,}\ee
where $\Sigma^{\rm HF}(1)$ denotes the Hartree-Fock shift.
This was numerically checked to be equivalent with the explicit
formula (\ref{ReSig}).

The bare nucleon-nucleon
interaction was approximated by a separable ansatz
\be\label{V_sep}
V_\alpha^{LL'}(p,p') = \sum_{i,j=1}^{\rm rank}w_{\alpha i}^{L}(p)
\lambda_{\alpha ij}w_{\alpha j}^{L'}(p')
\mbox{ .}\ee
The T-matrix then can be given algebraically as
\be\label{Tmat_sep}
T^{LL'}_\alpha(p,p',P,E) = \sum_{ijk}
w_{\alpha i}^L(p)
[1-J_\alpha(P,E)]_{ij}^{-1}
\lambda_{\alpha jk}w^{L'}_{\alpha k}(p')
\mbox{ ,}\ee
\be\label{jalpha}
J_\alpha(P,E)_{ij} = \int\frac{\dn^3 {\bf p}}{(2\pi)^3}\sum_{n L} w_{\alpha 
n}^L\lambda_{\alpha in}w_{\alpha j}^L(p)
\frac{<1-f({\bf P}/2+{\bf p})-f({\bf P}/2-{\bf p})>}
{E-\epsilon(P/2+ p)-\epsilon(P/2- p)}
\mbox{ ,}\ee
where $<...>$ denotes the usual angle averaging in the Pauli operator.
Having the selfenergy at our disposal
the spectral function follows from equation (\ref{A}).
The quasiparticle energies $\epsilon_1$ as defined by (\ref{eps}) were
determined in Hartree-Fock approximation
\be
\Sigma^{\rm HF}(1) =  \frac{1}{\Omega}\sum_{2}f(\epsilon_2)V_{\rm ex}(1,2,1,2)
\mbox{ .}\ee
With the T-matrix (\ref{Tmat_sep}) one is able to calculate the critical
temperature
for superfluidity using the Thouless-criterion \cite{Thouless}.
It has been demonstrated in ref.\ \cite{Thouless} that this
coincides with the
critical temperature found in BCS-theory.
For the separable
ansatz (\ref{V_sep}) it reads
\be\label{tc}
\det[1-J_\alpha(P=0,E=2\mu,T=T_c)]_{ij}=0
\mbox{ .}\ee
The Thouless-criterion for nuclear matter has already been evaluated in ref.\
\cite{AlmNP}.
\section{Selfenergy in Brueckner theory}
Brueckner theory is only applicable to zero temperature systems. The
near equivalence to Green function results has been demonstrated in
this case as already pointed out above. A $T>0$
extension of Brueckner theory 
was formulated by Bloch and De Dominicis \cite{blo}. An obvious
difference between the two approaches (Brueckner and Green function) is
that
the former only includes particle ladders while the latter also includes
hole ladders in defining the effective interaction. 
It is of some interest to numerically study the effect of this
difference. We therefore define a Brueckner K-matrix as in eq.
(\ref{T}) with T replaced by K. The propagator (\ref{G2}) is however
modified by the replacement
\be\label{prop}
1-f(\omega)-f(\omega')\longrightarrow (1-f(\omega))(1-f(\omega'))
\mbox{ .}\ee
The selfenergy ($V_{\rm B}$) in Brueckner theory is given by \cite{Malfliet1}
\bea\label{VBK}
V_{\rm B}(1,\omega) & = & \frac{1}{\Omega}\sum_{2}\Big(f(\epsilon_2)
K_{\rm ex}(1212,\epsilon_2+\omega)\nonumber\\
&&+{}\frac{1}{\Omega}\sum_{3,4}
{| K_{\rm ex}(1234,\epsilon_3+\epsilon_4) |}^2
\frac{[ 1 - f(\epsilon_2) ] f(\epsilon_3) f(\epsilon_4)}
{\omega+\epsilon_2-\epsilon_3-\epsilon_4-i\eta}
\delta_{{\bf p}_1+{\bf p}_2,{\bf p}_3+{\bf p}_4}
\Big)
\mbox{ .}\eea
For the discussion below, let us denote the
two terms in eq.\ (\ref{VBK}) by $V_{\rm B}^1$ and $V_{\rm B}^2$.
The second term, $V_{\rm B}^2$, is often referred to as the
Brueckner second order rearrangement potential. At a temperature $T=0$ it is
zero for states above the Fermi-surface and negative for states below,
while the first term is positive for states above and zero below the
Fermi-surface. 
As explained below, the retarded selfenergy in (finite-temperature)
Brueckner theory, denoted by $\Sigma_{\rm B}$, is given by
\be\label{ISIGB}
\im\,\Sigma_{\rm B}(1,\omega) = \im\,V_{\rm B}^1(1,\omega) -
\im\,V_{\rm B}^2(1,\omega)
\mbox{ ,}\ee
\be\label{RSIGB}
\re\,\Sigma_{\rm B}(1,\omega) = \re\,V_{\rm B}(1,\omega) =
\re\,V_{\rm B}^1(1,\omega) 
- {\bf P}\int\frac{d\omega'}{ \pi}
\frac{\im\,V_{\rm B}^2(1,\omega')}{\omega-\omega'}
\mbox{ .}\ee
This quantity is to be compared to the Green function $\Sigma$ defined 
in the previous section.

While $\Sigma_{\rm B}$ is calculated numerically according to eqs.\
(\ref{VBK}), (\ref{ISIGB}), and (\ref{RSIGB}), rewriting these
expressions to parallel those in the last section would facilitate
comparisons between the two formalisms.  To this end, one can,
invoking unitarity on the imaginary part of the second term
of eq.\ (\ref{VBK}),
write $\im\,V_{\rm B}$ as
\cite{Malfliet1}
\be\label{ImVB}
\im\,V_{\rm B}(1,\omega)=\frac{1}{\Omega}\sum_{2}
[f(\epsilon_2)-\bar{g}(\epsilon_2,\omega)]
\im\,K_{\rm ex}(1212,\epsilon_2+\omega)
\mbox{ ,}\ee
with
\be\label{gbar}
\bar{g}(\epsilon_2,\omega)=\frac{[1-f(\epsilon_2)]g(\epsilon_2+\omega)}
{g(\epsilon_2+\omega)+1}=f(\epsilon_2)e^{-\beta(\omega-\mu)}
\mbox{ .}\ee

Notice that at $2\mu=\epsilon_2+\omega$ where $g$ has a singularity
$\bar{g}$ is finite. 
It was argued in ref. \cite{hsk93} that the Brueckner $V_{\rm B}$ should
be identified 
with the chronological potential $\Sigma^{\rm c}$ which is related to
$\Sigma$ \cite{dan90,hsk93} by
\be\label{ImSigc}
\im\,\Sigma^{\rm c}(1,\omega)=
\tanh(\frac{1}{2}\beta(\omega-\mu))\,\im\,\Sigma(1,\omega)
\mbox{ .}\ee
In our notations here, this says, for $T>0$, one can obtain
$\im\,\Sigma_{\rm B}$ from
\be\label{ImSigB}
\im\,\Sigma_{\rm B} = \frac{\im\,V_{\rm B}}
{\tanh(\frac{1}{2}\beta(\omega-\mu))}
\mbox{ .}\ee
Now eqs.\ (\ref{ImVB}) and (\ref{gbar}) also give a simple relation between
the two terms contributing to $\im\,V_{\rm B}(1,\omega)$: 
\be\label{V12}
\im\,V_{\rm B}^2(1,\omega)=-e^{-\beta(\omega-\mu)}
\im\,V_{\rm B}^1(1,\omega)
\mbox{ .}\ee
It is then obvious that the division by the tanh function, in obtaining 
$\im\,\Sigma_{\rm B}$ from $\im\,V_{\rm B}$, is equivalent to just switching
the sign of $\im\,V_{\rm B}^2(1,\omega)$:
\be\label{ImSigB_VB1}
\im\,\Sigma_{\rm B}(1,\omega)=(1+e^{-\beta(\omega-\mu)})
\im\,V_{\rm B}^1(1,\omega)
\mbox{ ,}\ee
as stated in eq.\ (\ref{ISIGB}) above.

An equivalent way of expressing the foregoing correspondence between 
$\im\,V_{\rm B}(1,\omega)$ and Green function quantities is identifying
$2i\,\im\,V_{\rm B}^1$ and $2i\,\im\,V_{\rm B}^2$ with, respectively,
$\Sigma^<$ and $\Sigma^>$, the nonequilibrium extensions of which govern 
the loss and gain collision terms in a transport equation \cite{Botermans1}.
The on-shell version of this correspondence was pointed out by
Cugnon et al \cite{Cugnon1}.
\section{Results for the T-matrix (K-matrix) and the nucleon selfenergy}
In this exploratory calculation we used a rank one separable approximation
(Yamaguchi potential \cite{Yamaguchi}) as well as a rank two
parametrization of Mongan \cite{Mongan}.
The formfactors of the barely attractive Yamaguchi potential are
of the following form
\be\label{formf}
w_{\alpha}(p)=\frac{\lambda_{\alpha}}{p^2+\gamma^2}
\mbox{ ,}\ee
where the coupling constant and the effective range are given by
\bea
\lambda_{\alpha} & = & \left\{\begin{array}{rl}
12.3178\;(\MeV\,\fm^{-1})^{\frac{1}{2}} & \alpha={}^1{\rm S}_0\\
14.6988\;(\MeV\,\fm^{-1})^{\frac{1}{2}} & \alpha={}^3{\rm S}_1\end{array}
\right.\mbox{ ,}\non\\
\gamma & = & 1.4488\;\fm^{-1}
\mbox{ .}\eea
The formfactors for the rank two Mongan potential, which contains in addition
to a long-range attractive a short-range repulsive term,
have the same form as given in
eq.\ (\ref{formf}). The corresponding parametrization is given in
ref.\ \cite{Mongan}.
For numerical convenience we restricted
only to S-waves ($^1{\rm S}_0,{}^3{\rm S}_1$) which give the dominant
contribution to the T-matrix at low energies.

\begin{figure}
\centerline{\psfig{figure=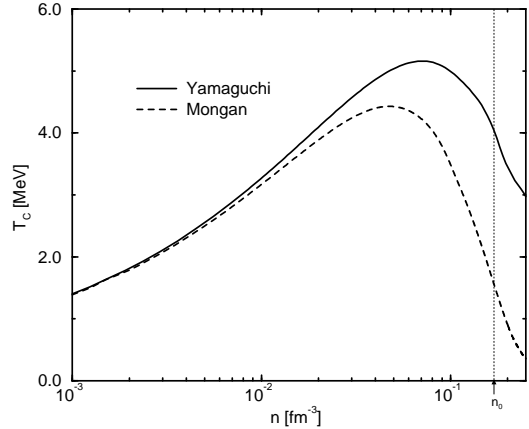,width=8cm,height=6.5cm}}
\caption{The critical temperature for superfluidity in symmetric nuclear matter
($^3{\rm S}_1$-channel) according to eq.\ (\protect\ref{tc}) as a function
of density for the Yamaguchi and the Mongan potential. The saturation density 
$n_0$ is indicated.}
\label{f_Tc}
\end{figure}

As we are concerned within this paper with the modification of the
selfenergy near the
critical temperature for the superfluid phase transition in
symmetric nuclear matter,
we first study the onset of superfluidity in the temperature-density plane
of nuclear matter.
In Fig.\ \ref{f_Tc} the critical temperature for superfluidity (\ref{tc})
($\alpha={}^3{\rm S}_1$)
is given as a function of the density. The solid curve
refers to the Yamaguchi potential whereas the dashed curve is for the Mongan
interaction.
Due to the repulsive component present in the Mongan interaction
the critical temperature is reduced compared to the barely attractive Yamaguchi
potential. These curves are in qualitative agreement with the calculations
done in ref.\ \cite{AlmNP}. In the following calculations we fix the density to
$n=n_0$ (dotted line in Fig.\ \ref{f_Tc} and vary the temperature.

\begin{figure}
\centerline{\psfig{figure=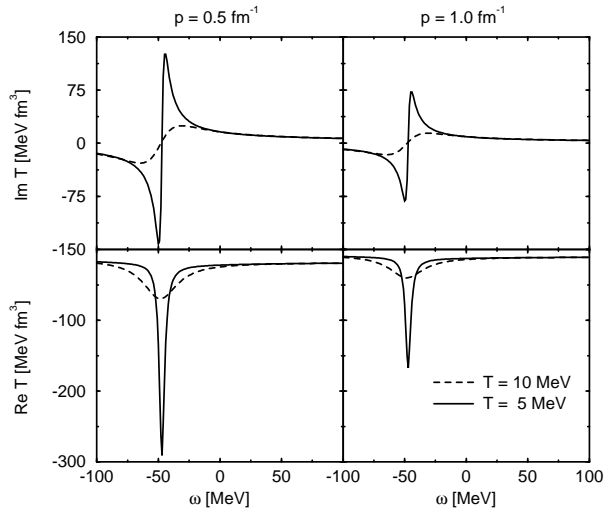,width=8cm,height=6.5cm}}
\caption{Real and imaginary part of the diagonal elements of the thermodynamic
T-matrix (\protect\ref{Tmat_sep})
in the $^3{\rm S}_1$-channel as a function of energy for
temperatures far above and
close to the critical temerature $T_c=4.02\;\MeV$ and for zero total momentum
at the saturation density. The relative momenta were set
to $p=0.5\mbox{ and }1.0\;\fm^{-1}$.}
\label{f_TMat}
\end{figure}

The key quantity for the calculation of the selfenergy within the
ladder-approximation is the thermodynamic T-matrix.
In Fig.\ \ref{f_TMat} the imaginary and the real part of the 
diagonal matrix elements of the thermodynamic T-matrix
(triplett-channel) of the Green function theory (\ref{Tmat_sep})
are given as a function of the energy argument
at fixed relative momentum $p$ and total momentum
$P=0$ as well as at a fixed chemical potential $\mu$.
The imaginary part shows a zero, which is located at $\omega=2 \mu$
independent on the temperature.
This is due to the fact that the imaginary part of the T-matrix
(\ref{Tmat_sep},\ref{jalpha}) is
proportional to
\be\label{pauli}
1-2f(\epsilon(p))=g^{-1}(2\epsilon(p))f(\epsilon(p))f(\epsilon(p))
\mbox{ ,}\ee
which is obviously zero at $2\epsilon(p)=2\mu$ for any temperature.
With decreasing temperature the slope of the imaginary part of the T-matrix
at $\omega=2\mu$ increases;
at temperatures near $T_c$ ($T_c=4.02\;\MeV$ in our case)
it has a characteristic principal
value structure.
The real part develops the corresponding peak at  $\omega=2\mu$
while approaching the critical temperature
from above. Finally at $T=T_c$ the real part of the T-matrix diverges
in accordance with the Thouless criterion  \cite{Thouless}.

The results of Fig.\ \ref{f_TMat} can be compared with the $T=0$ results
by Ramos et al. \cite{Ramos}. They also find a pairing instability
at $T=0$ and
$\omega=2\epsilon_{\rm F}$ which is most pronounced for small total momenta.
The difference compared to our results lies in the fact that although
their imaginary part goes to zero at $\omega=2\epsilon_{\rm F}$ as well,
it does not change sign at this energy as it does in our calculations.
This is due to the fact that in ref.\ \cite{Ramos} the Galitski-form
of the two-particle propagator in the kernel of the T-matrix  is used
\cite{Fetter} whereas we use the Kadanoff-Baym-form \cite{Kadanoff},
which results in the retarded T-matrix instead of the chronological one
used in ref.\ \cite{Ramos}.
In the limit $T\rightarrow 0$ both differ only in the sign of the
imaginary part for
$\omega>2\epsilon_{\rm F}$. Consequently both definitions differ also in the
sign of the corresponding selfenergies at $T=0$.
At finite $T$ the retarded selfenergy discussed so far and the
chronological selfenergy as derived from the Feynman-Galitski T-matrix
\cite{dan90} are related by eq.\ (\ref{ImSigc}).
\begin{figure}
\vspace*{0.5cm}
\centerline{\psfig{figure=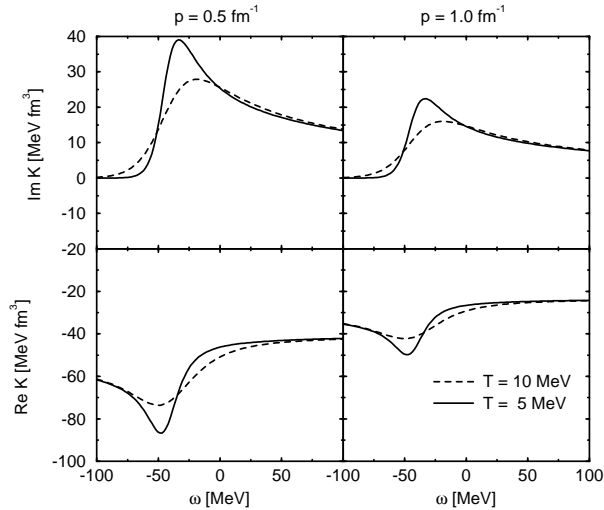,width=8cm,height=6.5cm}}
\caption{Real and imaginary part of the diagonal elements of the
thermodynamic K-matrix ($^3{\rm S}_1$) from Brueckner theory
plotted for the same parameters as in Fig.\ \protect\ref{f_TMat}.}
\label{f_KMat}
\end{figure}
However, with this difference in mind we can conclude that the pairing
instability at $T=T_c$ is in qualitative agreement with the
corresponding instability at $T=0$ observed in ref.\ \cite{Ramos}.
However, a proper treatment of the pairing correlations below $T_c$
demands the inclusion of a finite gap in the single particle
propagators. This has to be determined consistently from a combination
of the BCS theory with the T-matrix approximation. Such a treatment
known as quasiparticle-RPA \cite{rs} has recently been applied to
the one-dimensional Fermi gas \cite{Schuck}.

To compare with the Brueckner theory the respective key quantity is the
K-matrix.
In Fig.\ \ref{f_KMat} the K-matrix elements of the Brueckner theory
are given for the same set of parameters.
The imaginary part of the K-matrix does not change sign in contrast to the
T-matrix elements (Fig.\ \ref{f_TMat}).
This is due to the fact that in this case
the Pauli-blocking (\ref{prop}) is positive in contrast
to the Pauli-blocking term (\ref{pauli}), which changes sign at
$2\epsilon(p)=2\mu$. For low temperatures the imaginary part is
effectively zero for
energies below this particular energy. For energies
above this value a pronounced
maximum develops. This gives the corresponding maximum in the real part
of the K-matrix.
However, no critical temperature can be found where the K-matrix
diverges as found for the T-matrix (Thouless-criterion). Thus, the different
Pauli operator (\ref{prop}) in the K-matrix leads to considerable
deviations from the T-matrix case in particular in the limit
of low temperatures.

\begin{figure}
\centerline{\psfig{figure=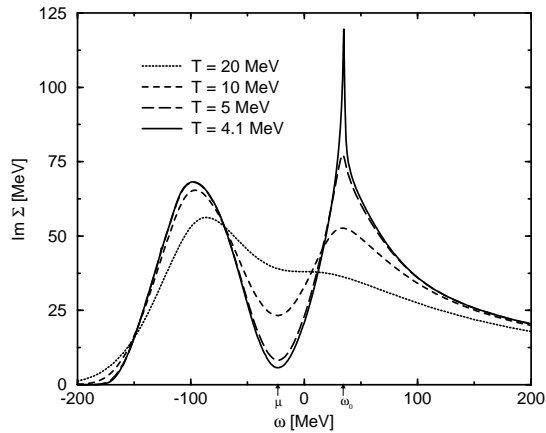,width=8cm,height=6.5cm}}
\caption{Imaginary part of the nucleon selfenergy (\protect\ref{ImSig})
at the saturation density $n_0$ as a function of energy for momentum $p_1=0$.
The selfenergy is given for several temperatures above the critical
temperature for superfluidity ($T_c=4.02\;\MeV$). The chemical potential
$\mu=-23.5\;\MeV$ ($T=4.1\;\MeV$) and the location of the pairing peak at
$\omega_0=2\mu-\epsilon_1=35.6\;\MeV$ are indicated.}
\label{f_ImSig_Tc}
\end{figure}

\begin{figure}
\centerline{\psfig{figure=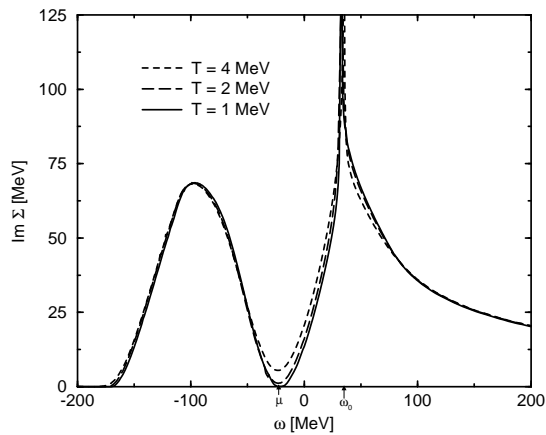,width=8cm,height=6.5cm}}
\caption{The same quantity as in Fig.\ \protect\ref{f_ImSig_Tc} for
temperatures
below the critical temperature $T_c=4.02\;\MeV$.
Again the chemical potential $\mu=-22.2\;\MeV$ ($T=1\;\MeV$) is indicated.}
\label{f_ImSig_T0}
\end{figure}

In order to calculate the spectral function it is necessary to evaluate the
off-shell selfenergy, which itself has some interesting features.
In Fig.\ \ref{f_ImSig_Tc} the imaginary part of the retarded selfenergy
(\ref{ImSig}) is given as a
function of $\omega$ at $p_1=0$ and $n=n_0$. The results are plotted
for various temperatures.
At the highest temperature $T=20\;\MeV$
(long-dashed curve) the selfenergy is rather smooth. It gives a non-zero
contribution for energies $\omega>-200\;\MeV$, develops a maximum near
$\omega=-100\;\MeV$ and then starts decreasing slowly.
In the $T=10\;\MeV$ case we find two additional extrema: a minimum at
the the chemical potential $\omega=\mu\approx -23\;\MeV$ and a second maximum
at $\omega_0=2\mu-\epsilon(p_1)\approx 35\;\MeV$.
Decreasing the temperature further this behaviour gets still more
pronounced. At the minimum the value of the imaginary part is
drastically reduced; however it still has a finite value and will reach
zero only in the limit $T\rightarrow 0$ (see discussion of
Fig.\ \ref{f_ImSig_T0}).
The second maximum gets more pronounced as well. At a temperature
$T=4.1\;\MeV$, close to the critical temperature $T_c=4.02\;\MeV$,
one observes a pronounced peak.
  
This particular behaviour can be traced back to the behaviour of
the T-matrix (\ref{Tmat_sep}) at low temperatures.
It has been demonstrated in ref.\ \cite{AlmPRC} that $\im\,T$ has a zero
at the particular energy value $z=2 \mu$ for pairs with zero total
momentum (compare also Fig.\ \ref{f_TMat}).
This compensates for the Bose singularity in the imaginary part of the
selfenergy (\ref{ImSig})  rendering Im $\Sigma$ at this
energy finite.
However, in addition a critical temperature can be found such that the
real part of the T-matrix becomes singular at the same energy
(Fig.\ \ref{f_TMat}).
As has been shown in Ref. \cite{AlmPRC} this critical temperature
coincides with the one for the superfluid phase transition in agreement
with the BCS theory. In fact, the singularity in the T-matrix is
nothing but the wellknown Thouless criterion \cite{Thouless} for superfluidity.
If this critical temperature is reached the above mentioned compensation 
does no longer hold (see also ref.\ \cite{AlmPRC}) 
 and the singularity in the T-matrix leads to a
corresponding singularity in the imaginary part of the selfenergy.
The singularity
is located at an energy $\omega_0=2\mu-\epsilon_1$. This is readily to
be seen if one restricts to the pole part of the T-matrix with total
momentum $P=0$. Then the integration over $p_2$
in (\ref{ImSig}) can be carried out directly yielding a $\delta$-peak in 
Im $\Sigma$ at the energy  $\omega_0$.
This shows that the singularity in the imaginary part of the selfenergy, which
occurs for $T\rightarrow T_c$ is a direct consequence of the pole in the
T-matrix at $T=T_c$, indicating the onset of superfluidity.

In Fig.\ \ref{f_ImSig_T0} we continued the evaluation of the imaginary
part of the
selfenergy for temperatures below $T_c$, disregarding for the moment
the pairing instability.
This allows us to demonstrate that for $T \rightarrow 0$
indeed the value of the imaginary part of $\Sigma$ at the minimum 
($\omega=\mu=-23\;\MeV$) approaches zero, in accordance with zero
temperature calculations of various authors \cite{Vonder1,Baldo}.
This zero of the imaginary part of the selfenergy at $\omega=\mu$
is a wellknown property at zero temperature leading to the fact that
particles at the Fermi surface have infinite lifetime, i.e. they are good 
quasiparticles.
According to the Hugenholtz-van-Hove theorem the chemical potential 
coincides with the binding energy per nucleon. The empirical value for
this quantity at saturation is $E_B^0/A=-16\;\MeV$.
The quadratic dependence of $\im\,\Sigma$ 
around $\omega=\mu$ at $T=0$ demonstrated in ref.\ \cite{Friman} is also
found in our numerical calculations.
With increasing temperature a non-zero value for $\im\,\Sigma$ is obtained
at $\omega=\mu$, however up to temperatures $T\le 20\;\MeV$
(see Fig.\ \ref{f_ImSig_Tc})
a pronounced minimum is reminiscent of the zero-temperature property.
At all temperatures $T<T_c$ the pairing instability shows up
at $\omega_0=2\mu-\epsilon_1=35\;\MeV$.
This pairing instability indicates the breakdown of the T-matrix approximation
in the vicinity of $\omega_0$ at temperatures below $T_c$.
A consistent treatment requires the inclusion of the BCS-gap for
temperatures below
$T_c$. Such a complicated calculation has not yet been carried out.
The $T=1\;\MeV$ curve in Fig.\ \ref{f_ImSig_T0} 
is in qualitative agreement with the calculation of ref.\ \cite{Koehler1}
except at energies around $\omega\approx 35\;\MeV$.
There we find the additional peak due to the pairing singularity in the
T-matrix discussed above.

\begin{figure}
\centerline{\psfig{figure=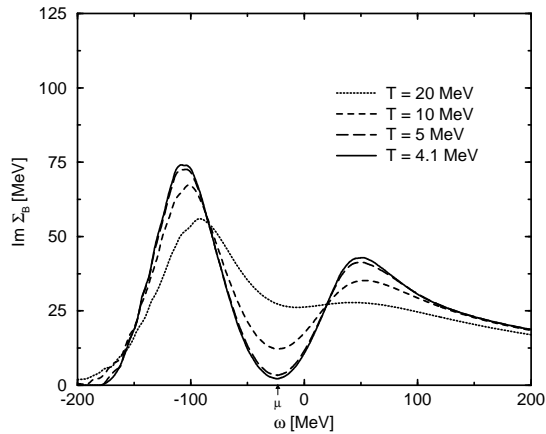,width=8cm,height=6.5cm}}
\caption{The imaginary part of the nucleon selfenergy
at the saturation density calculated in Brueckner theory
(eqs. (\protect\ref{ISIGB}))
for the same parameters as in Fig.\ \protect\ref{f_ImSig_Tc}.}
\label{f_ImSigB_Tc}
\end{figure}
In Fig.\ \ref{f_ImSigB_Tc} the imaginary part of the selfenergy
$\im\,\Sigma_{\rm B}$ (\ref{ISIGB}) is given as a
function of energy for the same parameters as in Fig.\ \ref{f_ImSig_Tc}.
For $T=20\;\MeV$
$\im\,\Sigma_{\rm B}$ is in qualitative agreement with the
corresponding curve in Fig.\ \ref{f_ImSig_Tc}.
With decreasing temperature the qualitative behaviour is
similar to the Green function
case given in Fig.\ \ref{f_ImSig_Tc} except in the energy range
around $\omega_0$.
Whereas a strong singularity is seen in Fig.\ \ref{f_ImSig_Tc} for
temperatures close to $T_c$
only a small maximum is found in the Brueckner case for the same temperature.
As discussed above this singularity is due to the pairing singularity
in the T-matrix
and occurs at the critical temperature defined by the Thouless criterion. 
This behaviour does not show up in the Brueckner theory.

\vspace*{1cm}
\begin{figure}
\centerline{\psfig{figure=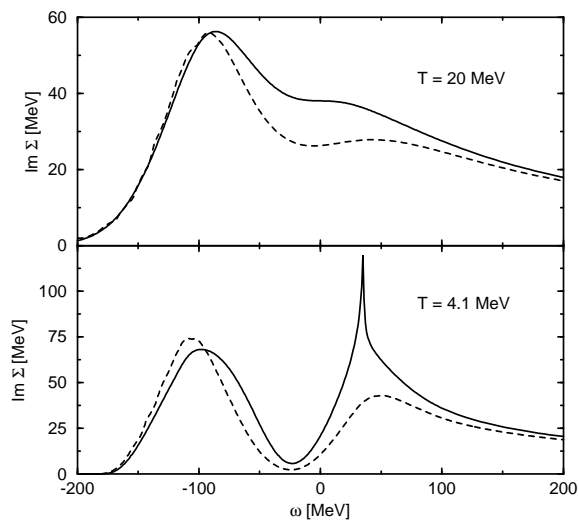,width=8cm,height=6.5cm}}
\caption{Direct comparison of the imaginary part of the selfenergy in Brueckner
theory (dashed curves) and Green function theory (solid curves) for
temperatures $T=20\mbox{ and }T=4.1\;\MeV$.}
\label{f_ImSig_GFBT}
\end{figure}
Fig.\ \ref{f_ImSig_GFBT} gives a direct comparison between the imaginary
parts of the selfenergy,
calculated in the T-matrix and the Brueckner approximations, respectively.
The upper curves show the case at $T=20\;\MeV$. While for energies below
$\omega=-100\;\MeV$ and above $\omega=100\;\MeV$ the two curves almost
coincide,
the imaginary part of the Brueckner selfenergy lies below the one from the
T-matrix calculation in the energy range between $-100\mbox{ and }100\;\MeV$.
An analogous result
has als been found in the zero-temperature case by K\"ohler \cite{Koehler1}.
In the $T=4.1\;\MeV$ case (slightly above $T_c$) we observe a similar
relation between the two
approximations. However, at energies around $\omega_0$ the Green
function selfenergy shows the pairing singularity, which is absent in
the Brueckner calculation.

\begin{figure}
\centerline{\psfig{figure=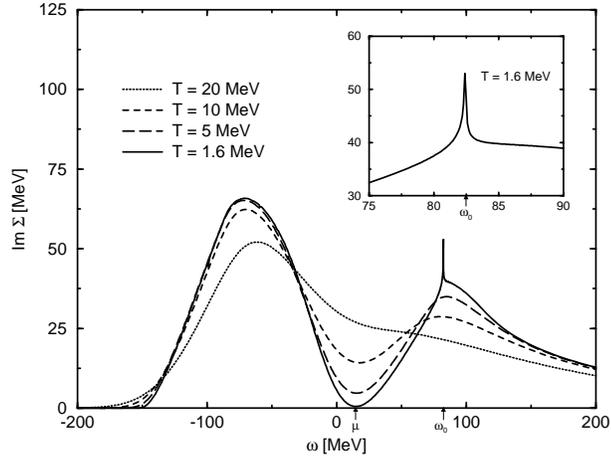,width=8cm,height=6.5cm}}
\caption{The imaginary part of the nucleon selfenergy at the saturation
density for the Mongan NN-interaction. The critical temperature in this case
has a smaller value ($T_c=1.58\;\MeV$) than for the Yamaguchi-potential.
Like in Fig.\ \protect\ref{f_ImSig_Tc}
the selfenergy is given for several temperatures above and close to
the critical temperature for superfluidity. The chemical potential
$\mu=14.68\;\MeV$ and the energy $\omega_0=82.46\;\MeV$ is indicated.
The inset shows the region around $\omega_0$ in detail.}
\label{f_ImSig_mong}
\end{figure}

In Fig.\ \ref{f_ImSig_mong} the same quantity like in Fig.\ \ref{f_ImSig_Tc}
is given using the Mongan potential
instead of the Yamaguchi potential used throughout the rest of the paper. The
purpose is to demonstrate the influence of the repulsive part of the
nucleon-nucleon interaction on the selfenergy.
\vspace*{1cm}
The repulsion leads to a lower critical temperature ($T_c=1.58\;\MeV$) as
compared
to the barely attractive case ($T_c=4.02\;\MeV$). The temperature
dependence of
$\im\,\Sigma$ does not change qualitatively compared to Fig.\ \ref{f_ImSig_Tc}.
Thus, we suppose that the behaviour
discussed above also holds for more realistic interactions, such as the
Paris-potential \cite{Plessas} as used e.g. in ref.\ \cite{AlmPRC}.

\begin{figure}
\centerline{\psfig{figure=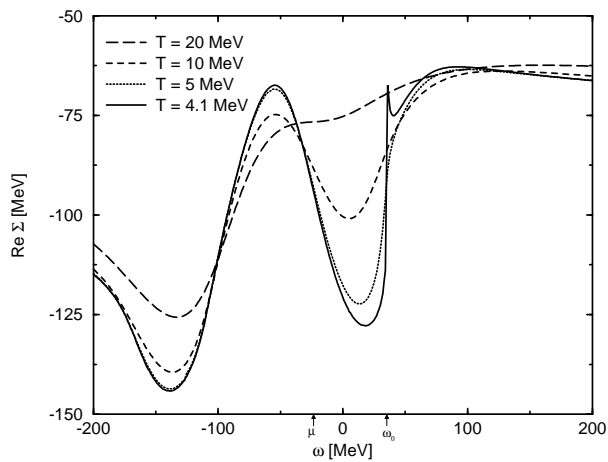,width=8cm,height=6.5cm}}
\caption{The real part of the off-shell nucleon selfenergy calculated from the
imaginary part according to eq.\ (\protect\ref{DisS})
as a function of energy for $n=n_0$ and the same temperatures as in
Fig.\ \protect\ref{f_ImSig_Tc}.}
\label{f_ReSig}
\end{figure}
In addition to the imaginary part the real part of the selfenergy
contains important
information, its on-shell part defining the optical potential for the nucleon
in nuclear matter.
In Fig.\ \ref{f_ReSig} the energy dependence of the real part of the
selfenergy as
calculated from eq.\ (\ref{DisS}) is shown for the same parameters as used
in Fig.\ \ref{f_ImSig_Tc}.
Again there is a relatively smooth behaviour for the $=20\;\MeV$ case. With
decreasingtemperature a second minimum is found for energies below $\omega_0$.
Please note that for $T=4.1\;\MeV$ (close to $T_c=4.02\;\MeV$)
a principle-value-like
behaviour around $\omega=\omega_0$ is found. This a direct consequence of the
corresponding pairing peak in the imaginary part of the selfenergy
at $\omega=\omega_0$ (see Fig.\ \ref{f_ImSig_Tc}).

\begin{figure}
\centerline{\psfig{figure=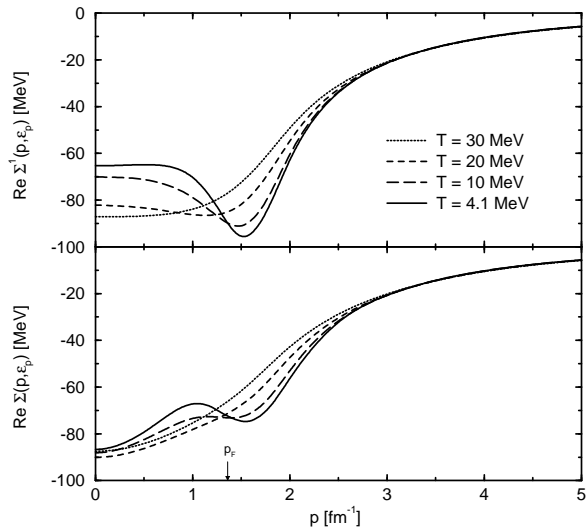,width=8cm,height=6.5cm}}
\caption{The real part of the on-shell nucleon selfenergy
as a function of the momentum $p$ for $n=n_0$ and different
temperatures according
to eq.\ (\protect\ref{ReSig}).
The upper figure was a calculation using only the first term of
eq.\ (\protect\ref{ReSig}) denoted as $\re\,\Sigma^1$.
The lower plot shows the result of the full expression.
The Fermi-momentum is indicated as $p_{\rm F}$.}
\label{f_ReSig_Tc}
\end{figure}

In Fig.\ \ref{f_ReSig_Tc} the real part of the on-shell self energy
i.e.\ the real part of the optical potential,
is given as a function of the momentum $p$.
The upper curve shows a calculation using only the first term in
eq.\ (\ref{ReSig}).
This is a standard approximation often used to calculate the optical
potential
in nuclear matter \cite{Schmidt,NYama,Lejeune}.
\vspace*{1cm}
With decreasing temperature one observes a pronounced minimum
near the Fermi energy,
which has first been observed in ref.\ \cite{Schmidt} and related
to the inclusion
of hole-hole scattering in the Pauli operator.
However, using the full expression of eq.\ (\ref{ReSig}) for the
evaluation of the
optical potential the behaviour at low temperatures is changed.
A particular structure is found around momenta
$p=p_{\rm F}$ which is enhanced with decreasing temperature. 
If one studies the behaviour of the real part
in detail, one finds, that the $^3{\rm S}_1$-channel of the T-matrix
is responsible for this anomalous behaviour.
We suppose, that the pairing peak in the imaginary part
$\im\,\Sigma(p,\omega)$ at temperatures
$T\rightarrow T_c$, present also at finite momenta $p$,
leads to a corresponding principal-value-like structure in the
real part of the optical potential. Restricting to the pole part of the
T-matrix this behaviour can be shown using the dispersion relation 
between the real and imaginary parts of the selfenergy.
Consequently for the on-shell selfenergy $\re\,\Sigma$ this leads to the wiggle
at $p=p_{\rm F}$.

\begin{figure}
\centerline{\psfig{figure=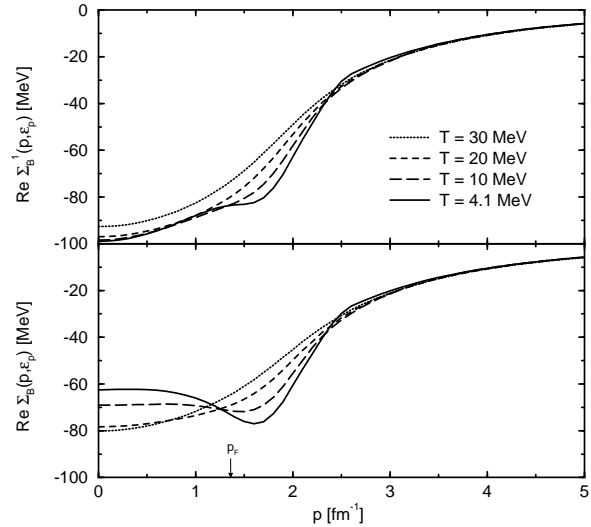,width=8cm,height=6.5cm}}
\caption{The real part of the on-shell nucleon selfenergy calculated with the
Brueckner K-matrix. The parameters are the same as in
Fig.\ \protect\ref{f_ReSig_Tc}. While the upper figure shows the results
only with the first term of eq.\ (\protect\ref{RSIGB})
denoted as $\re\,\Sigma_{\rm B}^1$ the
lower part shows the results of the full calculation.}
\label{f_ReSigB_Tc}
\end{figure}

Fig.\ \ref{f_ReSigB_Tc} shows the same quantity calculated from the
Brueckner theory.
The upper plot denoted as $\re\,\Sigma_{\rm B}^1$ shows the temparature
behaviour of
the first order Brueckner term (eq.\ (\ref{RSIGB})).
For the lowest
temperature ($T=4.1\;\MeV$) one observes a plateau-like behaviour around the
Fermi momentum. Using the full expressions (lower plot) the repulsive second
order contribution leads to a different behaviour at low temperatures, which
is characterized by a strong enhancement for low momenta and a corresponding
minimum for momenta $p>p_{\rm F}$.
\vspace*{1cm}

A particular behaviour of the optical potential for momenta around
$p=p_{\rm F}$
has also been found in refs.\cite{NYama,Ma1}.
In ref.\ \cite{Ma1} a plateau-like behaviour was related to the behaviour
of the effective mass at the Fermi surface.
In \cite{NYama} it was shown
that a non-monotonous behaviour around $p_{\rm F}$ (anomaly) was
entirely due to the
strong attraction
in the $^3{\rm S}_1-{}^3{\rm D}_1$ channel and that it was enhanced if
the density was decreased below $n=n_0$.
According to our understanding the anomaly observed in ref.\ \cite{NYama}
is probably as well due to the pairing instability \cite{Dickhoff}
discussed above.

\vspace*{0.5cm}
\begin{figure}
\centerline{\psfig{figure=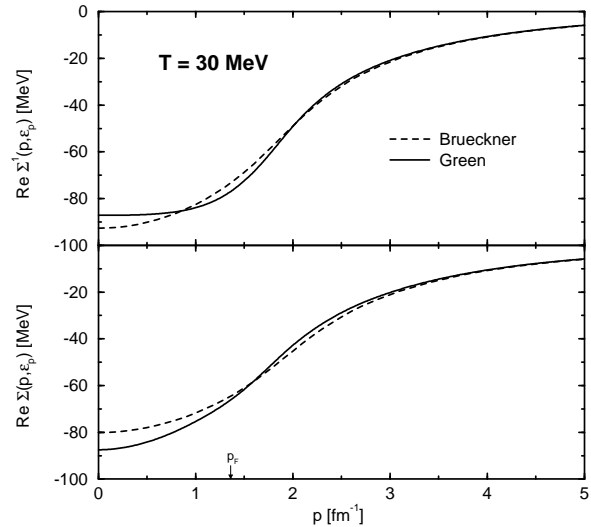,width=8cm,height=6.5cm}}
\caption{The real part of the on-shell nucleon selfenergy calculated in
T-matrix and Brueckner approximations, respectively, for temperature
$T=30\;\MeV$.}
\label{f_ReSig_T30}
\end{figure}
\begin{figure}
\centerline{\psfig{figure=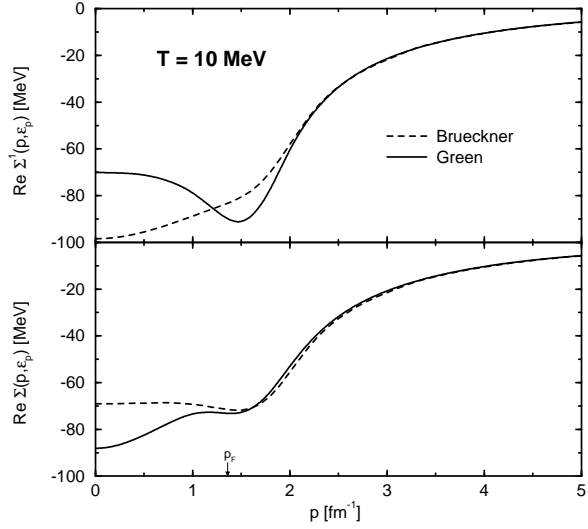,width=8cm,height=6.5cm}}
\caption{The same figure as Fig.\ \protect\ref{f_ReSig_T30}
for temperature $T=10\;\MeV$.}
\label{f_ReSig_T10}
\end{figure}
\begin{figure}
\centerline{\psfig{figure=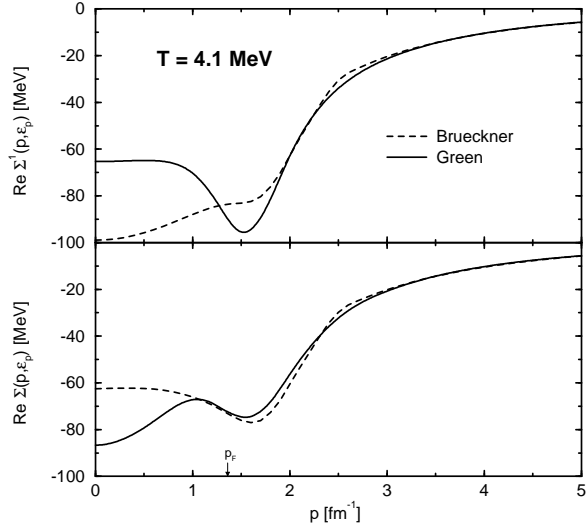,width=8cm,height=6.5cm}}
\caption{The same figure as Fig.\ \protect\ref{f_ReSig_T30}
for temperature $T=4.1\;\MeV$.}
\label{f_ReSig_T4}
\end{figure}

In Figs.\ \ref{f_ReSig_T30}-\ref{f_ReSig_T4} we give a direct comparison
between the Green function theory and
Brueckner theory with respect to the real part of the on-shell selfenergy for
different temperatures. The upper curves correspond to the first
term in eqs. (\ref{ReSig}) and (\ref{RSIGB}),
respectively.  As the form of these expressions coincides, the differences
between Green function and Brueckner theory
in this case are entirely due to the different Pauli operators in the T-matrix
and K-matrix, respectively (compare eq.\ (\ref{prop}).
The lower curves show the result of the full expressions
(\ref{ReSig}) and (\ref{RSIGB}).
In Fig.\ \ref{f_ReSig_T30} we give the results for temperature $T=30\;\MeV$.
The differences between
the two approaches are not very pronounced for this temperature, except for
low momenta. With respect to $\re\,\Sigma^1$ the Green-function-curve is
slightly enhanced
compared to the Brueckner-curve for momenta $p\simeq p_{\rm F}$ and
slightly reduced below. This behaviour is reversed for the lower curves.
\vspace*{0.5cm}
In Fig.\ \ref{f_ReSig_T10} ($T=10\;\MeV$) the differences between the two
theories are much more pronounced.
The different form of the Pauli operarator results in a non-monotonous
behaviour for the Green function curve showing a pronounced
minimum around $p=p_{\rm F}$. In contrast the Brueckner curve is
monotonously decreasing. A behaviour like this has first been
observed in ref.\ \cite{Schmidt}. For the full expressions the differences
between
both theories are less pronounced showing an enhancement of the Green function
curve with respect to the Brueckner curve for momenta $p<p_{\rm F}$.
\vspace*{0.5cm}
In Fig.\ \ref{f_ReSig_T4} the temperature ($T=4.1\;\MeV$) is close to the
critical temperature $T_c=4.02\;\MeV$.
One observes basically the same behaviour as in Fig.\ \ref{f_ReSig_T10},
although the differences are still more enhanced.

\begin{figure}
\centerline{\psfig{figure=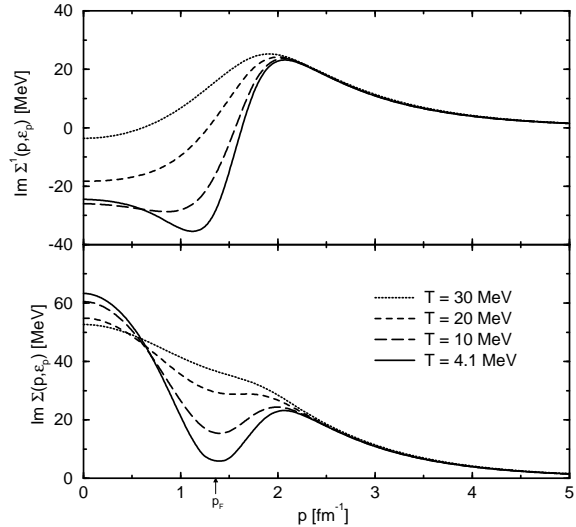,width=8cm,height=6.5cm}}
\caption{The imaginary part of the on-shell selfenergy as a function of $p$
at $n_0$ for the same
temperatures as in Fig.\ \protect\ref{f_ReSig_Tc},
according to eq.\ (\protect\ref{ImSig}).
The upper figure was a calculation using only the first term of
eq.\ (\protect\ref{ImSig}) denoted as $\im\,\Sigma^1$.
The lower plot shows the result of the full expression.
The Fermi-momentum is indicated as $p_{\rm F}$.}
\label{f_ImSig_GF}
\end{figure}

In Fig.\ \ref{f_ImSig_GF} the imaginary part of the on-shell selfenergy
(optical potential) is given as a function of momentum
for various temperatures. The upper graph shows
the contribution of the first term in eq.\ (\ref{ImSig}) only.
\vspace*{0.5cm}
One observes a pronounced temperature dependence
for momenta below $p=2\;\fm^{-1}$ leading
to a decrease with temperature towards a minimum near $p=p_{\rm F}$.
The lower graph
displays the full contribution for the imaginary part (\ref{ImSig}). Again,
one notes a strong temperature dependence below $p=2\;\fm^{-1}$. For momenta
around the Fermi momentum a pronounced minimum is exhibited with decreasing
temperature. The value of $\im\,\Sigma$ at $p=p_{\rm F}$ tends to zero
for $T\rightarrow 0$.
\vspace*{0.5cm}
\begin{figure}
\centerline{\psfig{figure=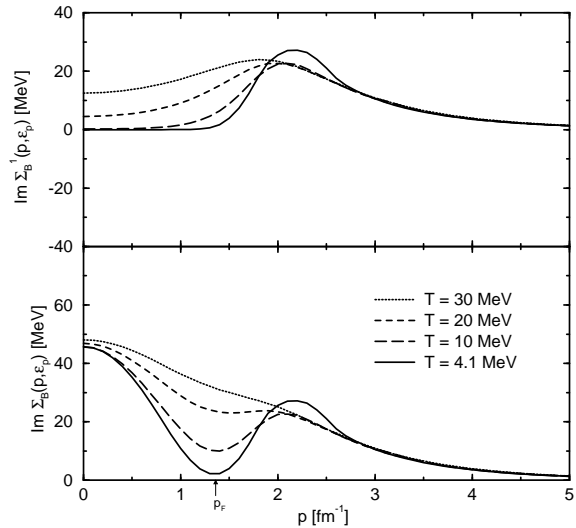,width=8cm,height=6.5cm}}
\caption{The imaginary part of the on-shell nucleon selfenergy calculated
with the
Brueckner K-matrix. The parameters are the same as in
Fig.\ \protect\ref{f_ImSig_GF}. The upper figure shows the results
only with the first term of eq.\ (\protect\ref{ISIGB})
denoted as $\im\,\Sigma_{\rm B}^1$.
The lower part shows the results of the full calculation
(\protect\ref{ISIGB}).}
\label{f_ImSig_BT}
\end{figure}
In Fig.\ \ref{f_ImSig_BT} the imaginary part of the on-shell selfenergy
calculated in
Brueckner theory $\im\,\Sigma_{\rm B}$ is given as a function of momentum for
the same parameters as in
Fig.\ \ref{f_ImSig_GF}. The upper plot displays the first term of
eq.\ (\ref{ISIGB}).
Again, we observe a strong decrease with temperature for momenta below
$p=2\;\fm^{-1}$ as in figure \ref{f_ImSig_GF}. However, due to the use of
the different Pauli-operator in the K-matrix $\im\,\Sigma_{\rm B}^1$ is
basically zero for
momenta $p<p_{\rm F}$ in the limit of low temperatures.
No minimum can be observed.
In the lower graph the full contribution of $\im\,\Sigma_{\rm B}$ (\ref{ISIGB})
is shown. One notes, that the temperature dependence for
momenta around $p_{\rm F}$ is in qualitative agreement with the
Green function result (lower curve of Fig.\ \ref{f_ImSig_GF}).
However, for momenta $p\leq 1\;\fm^{-1}$ the temperature behaviour is
reversed compared
to the Green function case.
Please note, that it is in the same momentum range, where one notes
pronounced deviations
in the real part of the optical potential at low temperatures (compare
lower part of
Fig.\ \ref{f_ReSig_T4}). The Brueckner results can be compared with the
calculation of
$\im\,V_{\rm B}(1,\epsilon_1)$ in ref.\ \cite{Grange1}.
Taking into account the relation between $\im\,\Sigma_{\rm B}$
and $\im\,V_{\rm B}$
(see sect. III) one notes the qualitative agreement of the two calculations.

Summarizing, one observes pronounced differences between the optical potential
at low temperatures calculated in Green-function-theory and Brueckner theory,
respectively. These
show up at momenta below the Fermi momentum. For the full expressions (lower
curves) these differences are due to higher order terms in the
Green-function-approximation,
not included in the second order Brueckner calculation (see also
ref.\ \cite{Malfliet1}).
\section{Spectral function and momentum distributions}
\begin{figure}
\centerline{\psfig{figure=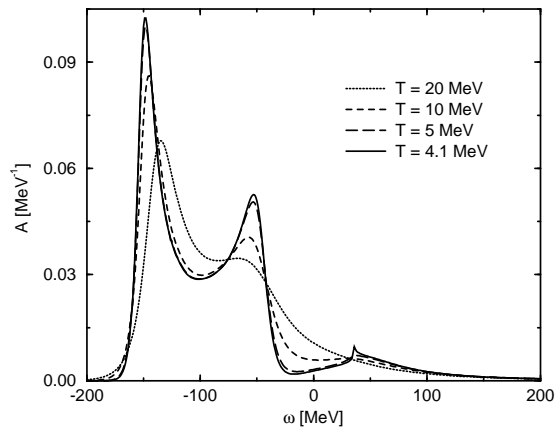,width=8cm,height=6.5cm}}
\caption{The nucleon spectral function at the saturation density as a function
of energy for momentum $p_1=0$.
The temperatures are the same as in Fig.\ \protect\ref{f_ImSig_Tc}.}
\label{f_A_Tc}
\end{figure}

From the off-shell selfenergy the nucleon spectral function is calculated using
eq.\ (\ref{A}).
In Fig.\ \ref{f_A_Tc} the nucleon spectral function  is plotted as a
function of energy $\omega$ for zero momentum $p_1=0$ at a density
$n=n_0$ and for the same temperatures as given  in Fig.\ \ref{f_ImSig_Tc}.
For $T=20\;\MeV$ one observes a quasiparticle peak at energies 
$\omega=- 150\;\MeV$ and a background contribution extending up to
energies $\omega\approx 200\;\MeV$.
With decreasing temperature the quasiparticle peak is slightly shifted
towards higher energies and its width is reduced.
This reduction is due to the fact that with decreasing
temperature the imaginary part reaches
zero at higher energies (compare Fig.\ \ref{f_ImSig_Tc}).
In addition a second maximum forms at lower energies.
This can be compared with the zero temperature results in ref.\
\cite{Koehler1} for the spectral function at $p=0$. 
Using Green function theory they also arrive at a spectral function with
two peaks of comparable size which are located
at approximately the same energies as given in Fig.\ \ref{f_A_Tc}.
in the $T=4.1\;\MeV$ case.

The temperature dependence of the spectral function is not as
drastic as one could expect from the change in the selfenergy with
temperature (Fig.\ \ref{f_ImSig_Tc}).
Near the critical temperature the tail of the spectral function at
higher energies shows additional smaller maxima, which result from the
pronounced structures in the real part of the selfenergy.
These in turn are due to the singular behaviour of the imaginary part
at $\omega=2\mu-\epsilon_1$.

\begin{figure}
\centerline{\psfig{figure=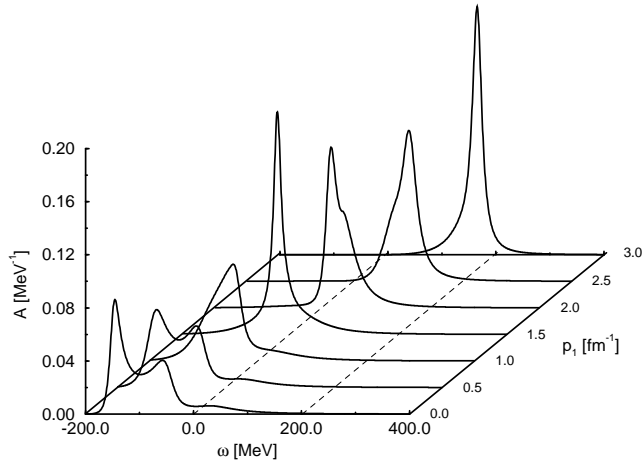,width=8cm,height=6.5cm}}
\caption{The nucleon spectral function at saturation density
and $T=10\;\MeV$ in the energy-momentum plane.}
\label{f_A_ep}
\end{figure}
In Fig.\ \ref{f_A_ep} the energy and momentum dependence of the spectral
function is given
at $n=n_0$ and $T=10\;\MeV$.
One observes that the double-peak structure found at $p=0$ vanishes with
increasing momentum. A single maximum remains for $p_1>0.7\;\fm^{-1}$
which can be identified with the quasiparticle peak.
The width of this peak is reduced at $p=p_{\rm F}$ due to the minimum
in $\im\,\Sigma$ at the chemical potential.
For larger momenta the peaks are broadened again until for
very high momenta the width is reduced again.
The latter behaviour is due to the fact that for very high momenta the
influence
of the medium represented by the selfenergy becomes negligible. Please note,
that for our choise of the nucleon-nucleon-interaction
there is no high-momentum tail of the
spectral function, because it is barely attractive.

\begin{figure}
\centerline{\psfig{figure=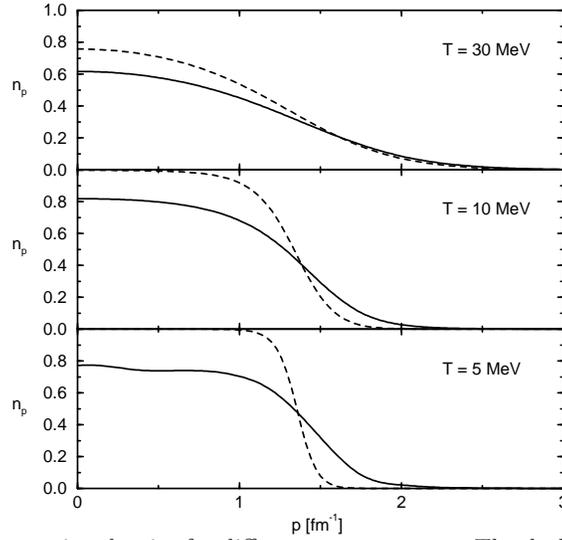,width=8cm,height=6.5cm}}
\caption{Occupation numbers at saturation density for different temperatures.
The dashed curves represent the uncorrelated occupation numbers, i.e. the
corresponding Fermi distribution.}
\label{f_occ_Tall}
\end{figure}

In order to demonstrate the influence of correlations on the nucleon occupation
numbers, the spectral function can be used to determine this quantity.
In Fig.\ \ref{f_occ_Tall} the temperature dependence of the nucleon
momentum distribution
(occupation numbers) $n(p)$ (\ref{occ_numb}) is given at a
fixed density $n=n_0$. 
The correlated occupation number (full line) is compared to the corresponding
Fermi distribution function (dashed line).
In the $T=5\;\MeV$ case we observe a strong depletion for momenta below the
Fermi momentum with a value of $n(p=0)=0.78$ compared to 1 for
the non-interacting case.
Above the Fermi surface we find a corresponding enhancement of the interacting
occupation numbers compared to the non-interacting up to about $p=2\;\fm^{-1}$.
For $T=10\;\MeV$ the depletion is less pronounced ($n(0)=0.82$).
This tendency towards
the non-interacting occupation numbers is continued is the case
of $T=30\;\MeV$.
With further increasing temperature the interacting response approaches the
non-interacting one.
Using the Brueckner K-matrix in ref.\ \cite{Koehler}
the finite temperature occupation numbers are evaluated
which are in reasonable aggreement with our results as well with the
calculations of ref.\ \cite{Grange1}.

\begin{figure}
\centerline{\psfig{figure=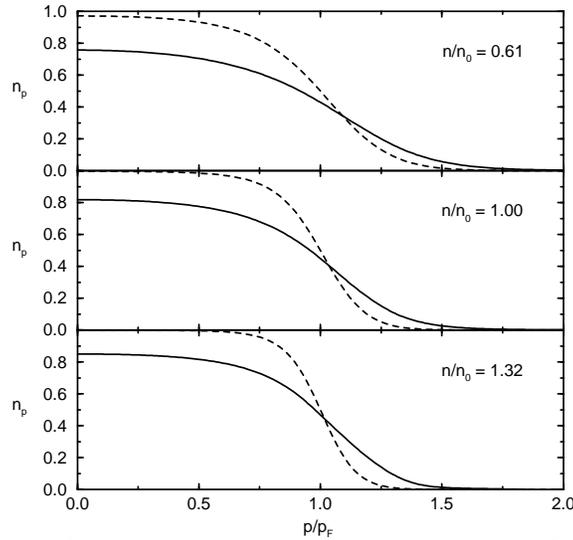,width=8cm,height=6.5cm}}
\caption{Occupation numbers at $T=10\;\MeV$ for density values around the
saturation density. Again the
dashed curves represent the non-interacting case.}
\label{f_occ_nall}
\end{figure}

In Fig.\ \ref{f_occ_nall} the density dependence of $n(p)$ at fixed $T$
($10\;\MeV$) is shown. For the
sake of a better comparability the curves are normalized to $p/p_{\rm F}$.
When going to densities above $n_0$ ($n/n_0=1.32$) one observes a lower
depletion ($n(p=0) = 0.85$) compared to $n(p=0) = 0.82$ at $n=n_0$.
\vspace*{0.5cm}
The same tendency has also been observed in ref.\ \cite{Koehler},
although at a higher temperature ($30\;\MeV$).
Going to lower densities ($n/n_0=0.61$) we find the astonishing result that the
depletion at low $p$ is further enhanced ($n(p=0) = 0.78$).
\vspace*{0.5cm}
On the other hand this corresponds to the zero temperature results of
ref.\ \cite{Ben1,Neck} indicating that the momentum distribution does not
approach
the non-interacting one in the limit of zero density. In ref.\ \cite{Neck} the
depletion at $T=0$ and $p=0$ stays rather constant for densities between
$n_0$ and $n_0/2$ at a value of $n(p=0) = 0.86$. The authors of
ref.\ \cite{Neck}
interpret this result as being due to the attractive part of the
nucleon-nucleon-interaction leading to the formation of bound pairs at low
densities.
\section{Summary and Conclusions}
Using the Matsubara Green function approach, selfconsistent expressions for the
nucleon selfenergy and the nucleon spectral function for nuclear matter
at finite temperature were derived.
The selfenergy and the nucleon spectral function at the saturation
density  were calculated in first iteration starting from the
quasiparticle spectral function.
The variation of these quantities with temperature was studied
for temperatures close to the critical temperature $T_c$ for the
superfluid phase transition in symmetric nuclear matter.
We found that approaching the critical temperature from above a
singularity develops in the imaginary part of the selfenergy.
It was shown that this
singularity is a direct consequence of a corresponding pole in the
T-matrix at the energy $E=2\mu$, which indicates
the onset of a superfluid phase
at $T=T_c$ \cite{Thouless}. Thus, the modification of the selfenergy near
$T_c$ can be understood as a precursor effect of the superfluid phase
transition in nuclear matter.
Another effect which was discussed as being related to the pairing instability
in the $^3{\rm S}_1-{}^3{\rm D}_1$-channel is the occurrence of a wiggle around
$p=p_{\rm F}$ in the real part of the on-shell selfenergy, especially
pronounced when approaching $T_c$ from above.
A similar effect although less pronounced
was also found in the Brueckner calculation.

The temperature dependence of the spectral function has been investigated for
temperatures above the critical temperature.  
Despite the strong modification of the selfenergy there is no
such drastic modification of the spectral function when approaching $T_c$
from above. This is consistent with the fact that below $T_c$ the condensate
part of the T matrix is proportional to the square of the gap and
consequently vanishes at the critical temperature.

The momentum dependence of the spectral function shows considerable deviations
from the quasiparticle behaviour at small momenta, whereas the quasiparticle
picture holds approximately for momenta around $p=p_{\rm F}$ as well as for
large $p$.

The occupation numbers were calculated from the spectral function at some
finite temperature and density.
It could be shown that with increasing temperature the non-interacting
occupation number is approached. The depletion of the occupation numbers
is enhanced with decreasing density at finite temperature.

At the end we would like to mention some open questions related
to our calculation of the nucleon spectral function:
The first question is related to the model interaction
we used in our exploratory
calculation of the selfenergy and spectral function.
For the energy and momentum range investigated in this paper
the important features
of the selfenergy obtained using the simple model interaction
of Yamaguchi type were also found using a rank two Mongan interaction.
It is supposed that these features remain of relevance also for
more realistic potentials \cite{Plessas}.

The second question is related to the problem of selfconsistency.
In principle, the spectral functions have to be iterated until
selfconsistency is reached. It remains to be seen, to what
extent the features
of the first iteration obtained in this calculation will also be found in a
fully selfconsistent calculation.

The third question concerns a consistent description of the 
system below $T_c$, where the consistent inclusion of a finite gap
is necessary for the evaluation of the spectral function.
In principle our calculation is restricted to temperatures
above the critical temperature for superfluidity.
The Thouless criterion indicates the instability of the
normal quasiparticle state with respect to the onset of superfluidity.
A consistent treatment below $T_c$ has to be based e.g. on a BCS
quasiparticle basis with a finite gap.
Up to now such a calculation has not been carried out for nuclear
matter. Instead, in most of the approaches at zero temperature
the implications of the pairing singularity for the selfenergy were
neglected. However, in a series of papers Dickhoff et al.
\cite{Vonder1}   stressed the need to properly take into account the
T matrix singularity, discussed above, which is present at
temperatures below $T_c$.

In conclusion we evaluated the nucleon self energy and spectral function for
finite temperature. We compared the calculations within
the Green function approach
with a finite temperature generalization of the Brueckner theory.
Special emphasis was put on the behaviour of these quantities near the critical
temperature for the onset of superfluidity in nuclear matter. Within the
Green function approach the pairing singularity in the T-matrix at the critical
temperature generates a corresponding singularity in the imaginary part of the
selfenergy. The non-monotonous behaviour (anomaly) of
the real part of the optical
potential for momenta $p\sim p_{\rm F}$ could also be related to the
pairing singularity.
The spectral function at finite temperature
shows a complex energy dependence, which cannot be generated from an
energy-independent width.

All the features discussed above cannot be incorperated into a
simple quasiparticle description.
Thus, the nucleon spectral function should be the appropriate quantity
for the description of hot and dense nuclear matter.

\end{document}